\documentclass[aps,prl,superscriptaddress,twocolumn,showpacs,amsmath,amssymb,floats]{revtex4}

\usepackage{txfonts}
\usepackage{amssymb}
\usepackage{graphicx}
\usepackage{hyperref}

\begin{document}

\title{Doping-controlled transition from excitonic insulator to semimetal in Ta$ _2 $NiSe$ _5 $}

\author{L. Chen}
\affiliation{International Center for Quantum Materials, School of Physics, Peking University, Beijing 100871, China}

\author{T. T. Han}
\affiliation{International Center for Quantum Materials, School of Physics, Peking University, Beijing 100871, China}

\author{C. Cai}
\affiliation{International Center for Quantum Materials, School of Physics, Peking University, Beijing 100871, China}

\author{Z. G. Wang}
\affiliation{International Center for Quantum Materials, School of Physics, Peking University, Beijing 100871, China}

\author{Y. D. Wang}
\affiliation{International Center for Quantum Materials, School of Physics, Peking University, Beijing 100871, China}

\author{Z. M. Xin}
\affiliation{International Center for Quantum Materials, School of Physics, Peking University, Beijing 100871, China}

\author{Y. Zhang}\email{yzhang85@pku.edu.cn}
\affiliation{International Center for Quantum Materials, School of Physics, Peking University, Beijing 100871, China}
\affiliation{Collaborative Innovation Center of Quantum Matter, Beijing 100871, China}

\date{\today}

\begin{abstract}

Excitonic insulator (EI) is an intriguing insulating phase of matter, where electrons and holes are bonded into pairs, so called excitons, and form a phase-coherent state via Bose-Einstein Condensation (BEC). Its theoretical concept has been proposed several decades ago, but the followed research is very limited, due to the rare occurrence of EI in natural materials and the lack of manipulating method of excitonic condensation. In this paper, we report the realization of a doping-controlled EI-to-semi-metal transition in Ta$ _2 $NiSe$ _5 $ using $in$-$situ$ potassium deposition. Combining with angle-resolved photoemission spectroscopy (ARPES), we delineate the evolution of electronic structure through the EI transition with unprecedented precision. The results not only show that Ta$ _2 $NiSe$ _5 $ (TNS) is an EI originated from a semi-metal non-interacting band structure, but also resolve two sequential transitions, which could be attributed to the phase-decoherence and pair-breaking respectively. Our results unveil the Bardeen-Cooper-Schrieffer (BCS)-BEC crossover behavior of TNS and demonstrate that its band structure and excitonic binding energy can be tuned precisely via alkali-metal deposition. This paves a way for investigations of BCS-BEC crossover phenomena, which could provide insights into the many-body physics in condensed matters and other many-body systems.

\end{abstract}

\maketitle

\begin{figure}[t]
	\includegraphics[width=7cm]{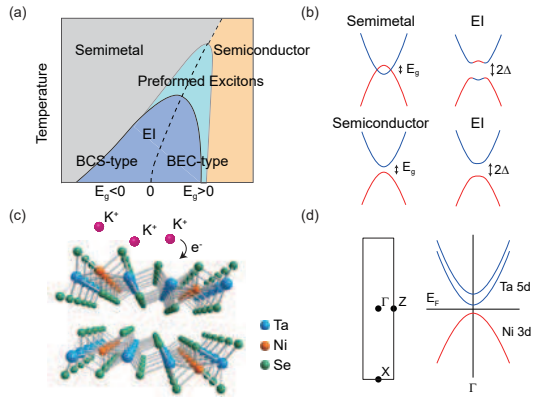}
	\caption{(a) The general phase diagram of EI predicted by theories \cite{gap5, gap6}. The dashed line indicates the crossover from BCS-type to BEC-type. (b) Schematic drawing of the band structures in a semi-metal, semiconductor, and EI. (c) Illustration of the lattice structure of TNS and the \emph{in-situ} potassium deposition. (d) Schematic drawing of the Brillouin zone of TNS and the predicted low-energy band structure. }\label{f1}
\end{figure}

Discoveries of exotic phases driven by many-body interactions have been motivating researchers in condensed matter physics for decades. In 1960s, in analogy with the Bardeen-Cooper-Schrieffer (BCS) theory, it was proposed that small gap semiconductors and small band-overlap semi-metals are unstable against the spontaneous formation of electron-hole pairs, namely excitons \cite{gap1, gap2, gap3}. Like the Cooper pairs condensation, the condensation of excitons also leads to a phase-coherent state, called excitonic insulator (EI). Depending on the way excitons form and condense, EIs can be classified into two categories [Fig.~\ref{f1}(a)] \cite{gap4, gap5, gap6}, the Bose-Einstein Condensation (BEC)-type EI where electrons and holes pair into excitons prior to the excitonic condensation, and the BCS-type EI where excitons form and condense simultaneously. More intriguingly, by tuning the band overlap ($E_g$), a BCS-type EI can evolve into a BEC-type EI smoothly and vice verse, which makes EI an ideal platform to study the many-body phenomena in the BCS-BEC crossover region. In semiconductors ($ E_g>0 $), the system favors a BEC-type condensation and the EI gap opens at the band-gap minima, while in semi-metals ($ E_g<0 $), the system favors a BCS-type condensation and the EI gap opens at the band crossings between the hole band (h-band) and electron band (e-band) [Fig.~\ref{f1}(b)]. In the BCS-BEC crossover region, the system cannot be described by either the BCS or BEC theoretical frameworks, many fascinating, yet not well understood phases such as the pseudogap phase, Fulde-Ferrell-Larkin-Ovchinnikov phase, etc. have been proposed \cite{cross1,cross2}.

Understanding EI especially in the BCS-BEC crossover region could provide insights into the many-body complexity of condensed matters and other many-body systems. However, even though the excitonic condensation theories have been established for decades, the followed experimental research is very limited. One reason is that the realization of EI is very difficult. The excitonic formation and condensation are highly sensitive to the electron-hole concentrations and screened Column interactions. So far, only the BCS-type EI has been observed in semiconducting bilayer structure \cite{gap7}. Many EI candidate materials, such as 1T-TiSe$ _2 $, Tm(Se$ _{0.45} $Te$ _{0.55} $), and Ta$ _2 $NiSe$ _5 $ (TNS) have been proposed \cite{gap8, gap9, gap10, gap11,gap14,gap15,gap16,gap23}. Besides the lack of EI materials, the non-thermal tuning method that tunes the EI condensation in natural materials is also lacking. As a result, it has not yet been able to observe directly how electronic structure evolves through an EI transition. These obstacles impede our understanding of the excitonic condensation phenomena, leaving many questions unanswered \cite{gap17,gap18, gap19, gap20, gap21}.

Here, we studied the electronic structure of TNS using angle-resolved photoemission spectroscopy (ARPES). By $in$-$situ$ depositing potassium on the sample surface, we succeeded in tuning TNS continuously from a EI to a semi-metal, which enables us to delineate the detailed evolution of electronic structure through the EI-to-semi-metal transition with unprecedented precision. Our data not only confirm that TNS is an EI originated from a semi-metal non-interacting band structure, but more importantly, we also resolve two transitions, the excitonic formation and condensation, respectively, which manifests a BCS-BEC crossover behavior. Our result clarifies previous controversies and puts strong constraints on theories. It establishes TNS as an ideal material for studies of the BCS-BEC crossover phenomena and also demonstrates that the alkali-metal deposition is an effective method for tuning the excitonic condensation in EI materials.

High-quality single crystals of TNS were synthesized using a chemical vapor transport (CVT) method \cite{gap12}. The chemical stoichiometry of our samples was confirmed by Energy Dispersive Spectroscopy (EDS). ARPES measurements were performed at Peking University using a DA30L analyser and a helium discharging lamp. The photon energy is 21.2~eV. The overall energy resolution was $\sim$~8~meV and the angular resolution was $\sim$~0.3$ ^\circ $. The crystals were cleaved \emph{in-situ} and measured in vacuum with a base pressure better than 6~$ \times $~10$ ^{-11} $~mbar. The experimental temperature was set to 80~K in order to avoid the surface charging effect. The potassium deposition was performed \emph{in-situ} using a potassium dispenser [Fig.~\ref{f1}(c)]. For each deposition step, we only kept the potassium evaporator at the working current for few seconds to achieve a fine doping step, and then took ARPES spectra. We repeated the deposition-and-measurement circle several times. The depositing time is defined as the total time of the potassium deposition. The total K-coverage is estimated to be $\sim$0.4~ML by counting the Fermi surface volume. For each step, K-coverage increases by $\sim$0.008~ML.

\begin{figure*}[t]
	\includegraphics[width=15cm]{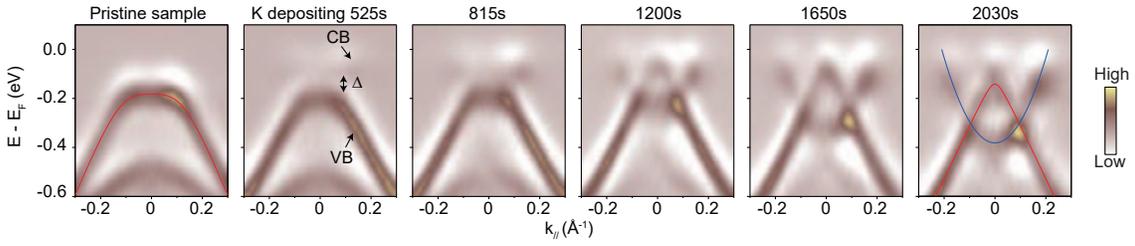}
	\caption{Evolution of the low-energy band structure in potassium-coated TNS. The photoemission spectra were taken along the $ \Gamma $-X direction at 80~K. The data were plotted using their second derivative images. Raw data can be found in Supplemental Material \cite{supplement}. The band dispersion in the pristine and heavily doped samples were fitted as described in Supplemental Material \cite{supplement}. }\label{f2}
\end{figure*}

TNS is a promising EI candidate \cite{gap10, gap11,gap14,gap15,gap16,gap23}. It has a quasi-one-dimensional crystal structure with one Ni-Se and two Ta-Se chains in each unit cell [Fig.~\ref{f1}(c)] . According to band calculations, the Ni-Se and Ta-Se chains contribute one h- and two e-bands, respectively, near the Fermi energy ($ E_F $) [Fig.~\ref{f1}(d)] \cite{gap23, gap12, gap13, gap22}. The h- and e- bands are either separated or overlapped by a small energy, which fits the theoretical requirement of EI. Fig.~\ref{f2} shows how the low-energy electronic structure of TNS evolves with potassium deposition. The pristine sample is an insulator. Its valence band (VB) is around 190 meV below $ E_F $. The VB band dispersion flattens around its band top, which deviates from a normal parabolic band dispersion. This flattening is consistent with previous ARPES results and is regarded as an evidence of excitonic condensation [Fig.~\ref{f1}(b)] \cite{gap8, gap11}. When doping with electrons, a conduction band (CB) emerges at $ E_F $. The energy gap between CB and VB shrinks quickly and eventually disappears at 1200~s. The band structure evolves into a semi-metal with one h-band and one e-band. No hybridization gap is observed between the h- and e-bands within our experimental resolution (Supplemental Material \cite{supplement}), indicating a weak inter-chain interaction between the electrons and holes. First, the e-band is quasi-one-dimensional and follows the crystal periodicity of TNS (Supplemental Material \cite{supplement}), which suggests that the e-band originates from TNS and cannot be attributed to the potassium-related two-dimensional electron gas. Second, most band calculations consider a three-band model when describing the low-energy band structure of TNS [Fig.~\ref{f1}(d)] \cite{gap22, gap23}. Here, the second e-band is found to be above $ E_F $ (Supplemental Material \cite{supplement}), and thus does not contribute to the phase transition. For simplicity, we only consider one e-band and one h-band in the following paper.

\begin{figure*}[t]
	\includegraphics[width=15cm]{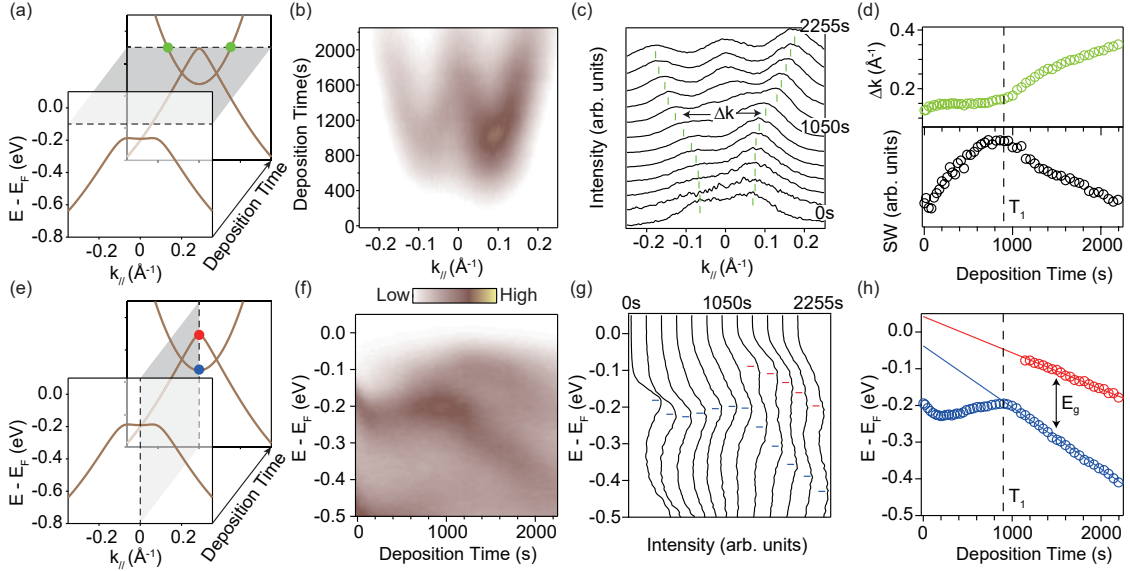}
	\caption{(a) Schematic drawing of the momentum-doping plane. The MDCs were taken at $ - $0.1~eV. (b) Merged image and (c) raw data of MDCs taken at different deposition times. The MDC peak positions are determined using Lorenz peak fitting and are marked using green ticks. The MDCs in (c) are intensity normalized to better visualize the peak positions. (d) Doping dependence of the peak separation ($ \Delta k $) (upper panel) and the integrated spectral weight (SW) (lower panel). (e), (f), and (g) are the same as (a), (b), and (c), but for the EDCs taken at the $ \Gamma $ point. The EDC peak positions are determined using Gaussian peak fitting and are marked using blue and red ticks. (h) Doping dependence of the h-band top (red circle) and the e-band bottom (blue circle) at the $ \Gamma $ point. The linear extrapolations of the band positions are shown by solid blue and red lines. The transition at 900~s (T$ _1 $) is illustrated by a dashed line. }\label{f3}
\end{figure*}

\begin{figure}[t]
\includegraphics[width=8.7cm]{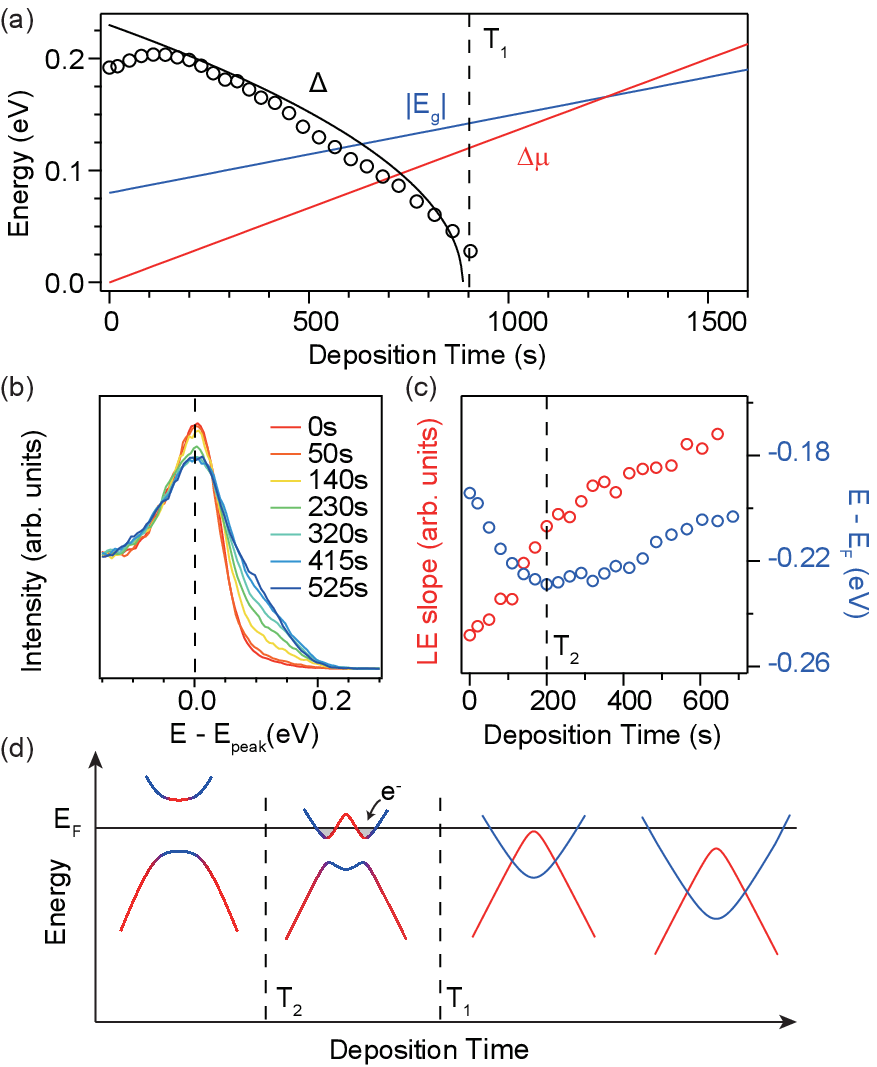}
\caption{(a) Doping dependence of the EI gap ($ \Delta $) (black circles), band overlap ($ |E_g| $) (blue solid line), and chemical potential shift ($ \Delta\mu $) (red solid line). The theoretical predicted gap equation ($ \Delta=\sqrt{\Delta_0(\Delta_0-2\Delta\mu)} $) is shown by a black sold line for comparison with $ \Delta_0 = 230 $~meV \cite{gap18}. (b) Doping dependence of EDCs taken at the $ \Gamma $ point. The peak positions are all aligned to better visualize the change of spectral line shape. (c) Doping dependence of the fitted leading edge (LE) slope (red circle) and the e-band bottom (blue circle). The transition at 200~s (T$ _2 $) is illustrated by a black dashed line. (d) Schematic drawing of the band evolution in potassium-coated TNS. The h- and e-bands are illustrated by the red and blue lines, respectively. }\label{f4}
\end{figure}

The evolution of low-energy electronic structure clearly shows an insulator-to-semi-metal transition in potassium-coated TNS. In comparison with the previous K-dosing work \cite{Karpes}, the clear doping effect and sharp ARPES spectrum suggest that our sample surface is free of any degradation or contamination during the potassium deposition, which could be attributed to our high-quality and well-controlled potassium dispenser. With such high-quality data, we could characterize the insulator-to-semi-metal transition quantitatively. We take the momentum distribution curve (MDC) at $ - $0.1~eV and plot its doping dependence in Fig.~\ref{f3}(a), \ref{f3}(b), \ref{f3}(c) and \ref{f3}(d). We choose $ - $0.1~eV instead of $ E_F $ to avoid the spectral weight suppression caused by the Luttinger-liquid behavior of TNS (Supplemental Material \cite{supplement}). The $ - $0.1~eV is also close to the band crossings of h- and e- bands where the energy gap opens (Fig.~\ref{f2}). The MDC peak positions represent the e-band positions, and the integrated spectral weight (SW) represents the electronic density of state at $ - $0.1~eV. As shown in Fig.~\ref{f3}(d), the doping dependences of both parameters clearly show a kink at around 900~s. The SW suppresses below 900~s, indicating an energy gap opening at $ - $0.1~eV. Correspondingly, the MDCs peak positions represent no longer the e-band positions, but the momenta of gap minima, when the energy gap opens. This explains the weak doping dependence of the MDC peak positions observed below 900~s. The transition is also obvious in the doping dependence of the energy distribution curve (EDC) taken at the zone center ($ \Gamma $) [Figs.~\ref{f3}(f), \ref{f3}(g) and \ref{f3}(h)]. In the heavily-doped semi-metallic region, the EDC consists of two peaks, representing the energy positions of the h-band top and the e-band bottom, respectively. With the decreasing of deposition time, while the h-band top cannot be tracked due to the spectral weight suppression near $ E_F $, the e-band bottom bends towards higher binding energy at 900~s, indicating an energy gap opening.

The alkali-metal deposition normally affects the electronic structure in two different ways. One is the carrier doping, which raises the system's chemical potential. The other one is the establishment of surface electric field, which changes the band overlap between CB and VB due to the Stark effect \cite{gap25, gap26, gap27, gap28}. Our data clearly show both effects. In the semi-metallic region, the band positions shift linearly towards higher binding energy with potassium deposition, indicating an increment of chemical potential. Meanwhile, the band overlap between h- and e-bands also increases linearly. We could then fit the band shifts in Fig.~\ref{f3}(h) and assume that this linear relationship could persist to the pristine sample, if there is no energy gap opening and the system remains semi-metalic. The non-interacting band structure of the pristine TNS could then be determined by the linear extrapolation [Fig.~\ref{f3}(h)]. The obtained non-interacting band structure of TNS is a semi-metal with a band overlap around 80~meV. Firstly, the h-band top and e-band bottom are nearly symmetric to $ E_F $, which fits the electric-neutrality requirement of TNS. Secondly, our observation of two well separated MDC peaks in both the lightly-doped and pristine samples is consistent with a gapped semi-metallic band structure, where two gap minima locate at the band crossings between the h- and e-bands [Fig.~\ref{f1}(b)]. Thirdly, the momentum separation between the two gap minima ($ \Delta k $) is around $ 0.12\;\AA^{-1} $ in the pristine sample, which is also consistent with the 80 meV band overlap (Supplemental Material \cite{supplement}). These consistencies prove the validity of our linear extrapolation and confirm that the non-interacting band structure of TNS is a semi-metal.

With the non-interacting h-band top ($ E_h $) and e-band bottom ($ E_e $) as well as the measured e-band bottom ($ E_e^{'} $), we can calculate the EI gap ($ \Delta $), using the equation \cite{gap6} $ \Delta=\sqrt{\left[(E_h+E_e)/2-E_e^{'}\right]^2-\left[ (E_h-E_e)/2\right] ^2} $. The band overlap is calculated using $ \left| E_g\right|=\left|E_h-E_e\right| $, and the shift of chemical potential is calculated using $ \Delta\mu=\left|(E_h+E_e)/2-(\left.E_h\right|_{0s}+\left.E_e\right|_{0s})/2\right| $. The results are plotted in Fig.~\ref{f4}(a). The EI gap in the pristine sample ($ \Delta_0 $) mirrors the excitonic binding energy of TNS. It is around 200~meV, which is consistent with the energy scale determined by optical spectroscopy experiments \cite{gap14, gap15}. The EI gap shrinks with potassium deposition and closes at around 900~s. At the gap closing transition (T$ _1 $), $ \left|E_g\right| $ is around 140 meV and $ \Delta\mu $ is around 120~meV. Besides T$ _1 $, the band evolution shows another abnormal behavior at around 200~s (T$ _2 $) [Fig.~\ref{f3}(h) and \ref{f4}(c)]. The rapid shift of band position below T$ _2 $ can be explained by a shift of chemical potential from the gap edge to the gap center. This indicates a depletion of free electrons in the system. Meanwhile, as shown in Fig.~\ref{f4}(b), the EDC peak narrows below T$ _2 $, which is analogy to the phase coherent behavior observed in high-$ T_c $ cuprates \cite{gap29}. We fitted the leading edge slope and plot its doping dependence in Fig.~\ref{f4}(c). The leading edge slope steepens at around 200~s, indicating an increment of phase coherence at T$ _2 $. The evolution of electronic structure in potassium-coated TNS is summarized in Fig.~\ref{f4}(d). When going from the semi-metallic phase to the EI phase, the EI gap first opens at the band crossings between the h- and e-bands at T$ _1 $. Then, at T$ _2 $, the chemical potential shifts from the gap edge to the gap center, and meanwhile, the spectra become phase-coherent.

Whether the insulating behavior of TNS originates from exactionic condensation or not is under debate. Some argue that its insulating property originates from a symmetry breaking of lattice instead of excitonic condensation \cite{gap24}. According to theoretical studies, both the carrier doping and band overlap are critical control-parameters for EIs \cite{gap2, gap18}. It has been proposed that the EI phase is unstable or energetically not favored when the band overlap exceeds the excitonic binding energy ($ E_g>\Delta_0 $) \cite{gap2} or the shift of chemical potential exceeds the half of the excitonic binding energy ($ \Delta\mu>\Delta_0/2 $) \cite{gap18}. Our results are quantitatively consistent with these proposals [Fig.~\ref{f4}(a)], suggesting that the observed gap opening is related to an excitonic formation. On the other hand, the energy gap opens at the band crossings instead of at $ E_F $. This excludes that the energy gap is induced by a charge-density-wave (CDW) instability driven by the Fermi surface instability. The energy scale of $ \Delta_0 $ is also too large to be explained by the symmetry breaking of lattice or the structure distortion \cite{gap30}. All these results suggest that TNS is an EI and undergoes an EI-to-semi-metal transition with electron doping.

Along temperature axis, TNS undergoes an semiconductor-to-insulator transition at 328~K \cite{gap12, gap13}. Here, we show that the non-interacting band structure of TNS is a semi-metal. While this clarifies previous controversy regarding to the non-interacting band structure of TNS \cite {gap10, gap11, gap14, gap15, gap16, gap24}, it also suggests that the semiconducting phase of TNS above 328~K is not a normal state but instead could be an exciton pre-pairing state \cite{gap10, gap11, gap16}. Under this scenario, excitons form already at a much higher temperature and the transition at 328~K could be attributed to an excitonic condensation. Along the doping axis, we resolved clearly two separated transitions. The T$ _1 $ at 900~s is characterized by a clear excitonic gap opening, which indicates a formation of excitons. At T$ _2 $, the spectra become phase-coherent indicating an excitonic condensation. Both the semi-metallic non-interacting band structure and the separated excitonic formation and condensation put TNS in the BCS-BEC crossover region. Meanwhile, the band overlap of TNS can be tuned by pressure or chemical substitution of Se by S \cite{gap13, gap31}. These make TNS an ideal material to achieve a continuous transition from a BCS-type EI to a BEC-type EI. It is then intriguing to study the BCS-BEC crossover phenomena in TNS. The results would establish a connection between the BCS and BEC theocratical frameworks and help us to understand the many-body phenomena in condensed matters and other systems.

Our detailed delineation of how electronic structure evolves across the EI transition puts strong constraints on theories. It has been proposed that a spin-polarized state would emerge in electron-doped EI \cite{gap18, gap19, gap20, gap21}. However, no band splitting is observed within our experimental resolution. Moreover, the large energy scale of the excitonic gap and the BCS-BEC crossover behavior observed in TNS cannot be explained using the weak-coupling theories of EI. Strong-coupling theories or the interplay between excitons and other degrees of freedomn, such as phonons, should be considered \cite{gap16, gap23}. Finally, we demonstrate that the alkali-metal deposition is an effective way to tune the EI transition. This method can be used extensively in transport, scanning tunneling microscopy, optical, and ARPES experiments to verify the existence of EI in other candidate materials such as 1T-TiSe$ _2 $ and Tm(Se$ _{0.45} $Te$ _{0.55} $). It also paves a way for the manipulation of excitonic-related phenomena in solids, and potentially helps to seek for excitonic superconductors.

This work is supported by the National Natural Science Foundation of China (Grant No. 11888101), by the National Key Research and Development Program of China (Grant No. 2018YFA0305602 and No. 2016YFA0301003), by the National Natural Science Foundation of China (Grant No. 91421107 and No. 11574004).

\end{document}